\documentclass[reprint,superscriptaddress,amsmath,amssymb,aps,prb,showkeys,showpacs]{revtex4-1}
\usepackage{graphicx}
\usepackage{color}

\def\ve{\varepsilon}

\def\w{\omega}

\def\dg{\dagger}

\newcommand{\ket}[1]{\mathinner{|{#1}\rangle}}
\newcommand{\bra}[1]{\mathinner{\langle{#1}|}}

\begin{document}
\title{Autler-Townes Doublet observation via a Cooper-Pair Beam Splitter}
\author{M. O. Assun\c{c}\~ao}
\affiliation{Instituto de F\'{\i}sica, Universidade Federal de Uberl\^andia, 38400-902 Uberl\^andia, MG, Brazil}
\affiliation{Curso de F\'{\i}sica, Universidade Federal de Jata\'i, 75801-615 Jata\'{\i}, GO, Brazil}
\author{G. S. Diniz}
\affiliation{Curso de F\'{\i}sica, Universidade Federal de Jata\'i, 75801-615 Jata\'{\i}, GO, Brazil}
\author{L. Sanz}
\affiliation{Instituto de F\'{\i}sica, Universidade Federal de Uberl\^andia, 38400-902 Uberl\^andia, MG, Brazil}
\author{F. M. Souza}
\affiliation{Instituto de F\'{\i}sica, Universidade Federal de Uberl\^andia, 38400-902 Uberl\^andia, MG, Brazil}
\email{fmsouza@ufu.br}
\date{\today}

\begin{abstract}

We present a \textit{proof-of-principle} of how electronic transport measurements permit the observation of the Autler-Townes doublet, an optical property of nanodevices. The quantum physical system consists of one optically pumped quantum dot, a second auxiliary quantum dot, and a superconductor lead which provides an effective coupling between the dots via crossed Andreev reflection. Electrodes, working as sources or drains, act as nonequilibrium electronic reservoirs. Our calculations of the photocurrent at both, transient and stationary regimes, obtained using a density matrix formalism for open quantum systems, shows signatures of the formation of the Autler-Townes doublet, caused by the interplay between the optical pumping and the crossed Andreev reflection.

\end{abstract}
\keywords{nonequilibrium quantum systems, quantum transport, quantum dots.}
\maketitle
\section{Introduction}
\label{sec:intro}
Since the seminal work of Loss and DiVincenzo on quantum computation~\cite{loss98},
semiconductor quantum dots have become an outstanding system for future development of
integrated photonic and electronic scalable devices~\cite{10.1063/1.5021345,PhysRevApplied.6.054013}.
Such an integration has fundamental importance on the development of quantum computers and quantum
internet~\cite{kimble08,PhysRevA.83.012303}. In the last case, a quantum network is required,
where fixed quantum nodes, defined by the quantum dots, exchange information through flying qubits,
those codified on the state of the photons~\cite{delteil2017,he2017}. Such an implementation is feasible
because of the entanglement between spin states on quantum dot and a single photon~\cite{greve2012,gao2012}.

Quantum dots under the action of laser fields have been intensively investigated in recent years~\cite{freitas17,majumdar12,obrien09}.
The optical response of such a level configuration is rich, ranging from the appearance of the so-called
Autler Townes doublet (ATD)~\cite{autler1955}, to robust states and tunneling induced
transparency (TIT)~\cite{xu2007,jundt2008,borges2010,PhysRevLett.107.163604,borges2012}.
Originally, ATD was first reported in a molecular system composed of gaseous carbonyl sulfide (OCS)
being excited by a \emph{rf} field and probed via a microwave field~\cite{autler1955}.
It occurs in a three-level system where a double transition is observed between dressed states by the presence of a radiation field~\cite{Scullybook}.
Since then, it has been observed in atoms~\cite{boller1991} and superconductor qubits~\cite{sillanpaa2009,novikov2013,peng2017}.
More recently, it was predicted that a nonlinearity in the current of a photodiode will appear when one of the
ATD splitted levels crosses the Fermi level of electronic reservoirs~\cite{assuncao2013}.

Extending the functionalities of quantum dots, they have also been used to create hybrid systems
composed of quantum dots and superconductors~\cite{PhysRevB.83.125421,PhysRevB.63.165314,franceschi2010,PhysRevB.92.054514,sherman2017,zhaoen2017}.
One device known as a Cooper pair beam splitter has been implemented
using different experimental setups~\cite{hofstetter2009,herrmann2010,PhysRevB.96.195409,herrmann2010}.
In these systems, a Cooper pair is split into two electrons going to different contacts
in a process known as \emph{crossed Andreev reflection} (CAR)~\cite{PhysRevB.59.3831,deutscher2000,Falci01,deutscher2002,PhysRevB.68.174504,PhysRevLett.93.197003,PhysRevLett.95.027002}. This effect attracted a great deal of attention in the last two decades, including works in the context of quantum dots~\cite{hiltscher2011,PhysRevB.91.235424,PhysRevLett.120.087701}, mainly due to possibility of spin-entangled electrons formation~\cite{PhysRevB.63.165314}.

Here, we present a \emph{proof-of-principle} of probing the formation of the ATD through quantum transport.
In the literature, the ATD is reported only through spectroscopic measurements~\cite{Gall11,bauer2009,Leeuwen95}.
As far as we are concerned, this is the \emph{first proposal}
to detect ATD relying on measurements of photoinduced current,
driven by optical pumping of quantum dots. The interplay between CAR and the electromagnetic
field can be mapped into a three-level system, that sustains an Autler-Townes doublet in a nonequilibrium regime.

This paper is organized as follows: in Secs.~\ref{sec:model} and \ref{closedsystem}, we present the effective model and a discussion of the physical
setup that allows the formation of an Autler-Townes doublet, along to the closed dynamics defined in the effective three-level quantum
systems. Section~\ref{sec:dynopen} presents the theoretical treatment for studying the action of reservoirs. In Sec.~\ref{sec:photocurrent}, we discuss
the behavior of photocurrent, which has signatures of ATD and finally, in Sec.~\ref{sec:summary} contains the final remarks.

\section{Physical system and the Effective Model}
\label{sec:model}
\begin{figure}[tb]
\centering\includegraphics[width=0.9\linewidth]{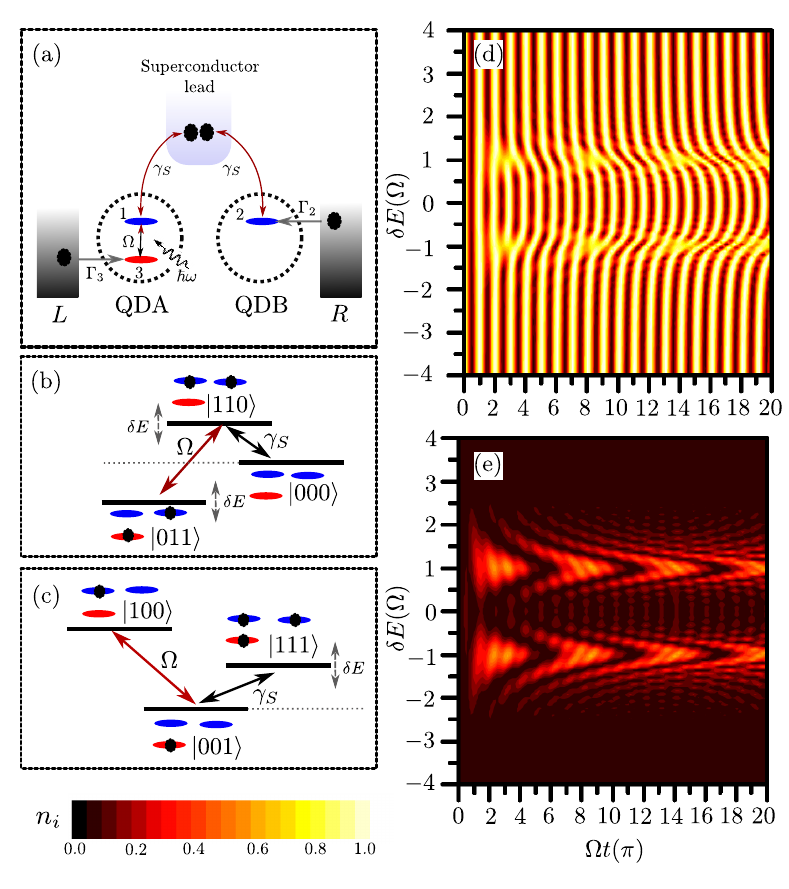}
\caption{(a) Illustration of the experimental setup: a laser pumping, with strength $\Omega$ and frequency $\omega$, promotes electrons from valence ($3$) to conduction ($1$) band on QDA. CAR creates an effective coupling,
quantified by $\gamma_{S}$, between the CBs of QDA (1) and QDB ($2$). Reservoirs $L$ and $R$ act as sources,
as shown here, or drains, through incoherent tunneling rates $\Gamma_i$.
(b)-(c) Schematic representation of energy parameters and couplings of the effective three-level systems:
(b) $\Lambda$ configuration: for states with even number of particles, and (c) $V$ configuration:
for states with odd number of particles. (d)-(e) Closed dynamics of the quantum system ($\Gamma_2=\Gamma_3=0$),
as function of time and $\delta E=\varepsilon_2-\varepsilon_1$, considering the initial
state $\ket{001}$ ($V$ configuration): (d) occupation of VB of QDA, $n_3$ and (e) occupation of CB of QDB, $n_2$.
Physical parameters are set as $\varepsilon_1=0$, $\omega=\varepsilon_1-\varepsilon_3$, and $\gamma_{S}=\Omega/4$.}
\label{fig:figure1}
\end{figure}

A schematic illustration of the physical setup is shown in Fig.~\ref{fig:figure1}(a),
consisting of two quantum dots: quantum dot A (QDA) and quantum dot B (QDB),
which are indirectly coupled to each other via a superconductor lead.
This geometry was proposed in Ref.~[\onlinecite{hiltscher2011}] in order to investigate adiabatic
pumping in a Cooper pair beam splitter (CPBS).
A CPBS creates or annihilates a pair of electrons with opposite spins on the conduction band (CB) of QDA and QDB in a process known as \emph{Crossed Andreev Reflection} (CAR)~\cite{hiltscher2011,PhysRevB.91.235424}.
We assume that the superconductor gap is large enough,
thus forbidding tunneling of single particles to the superconductor~\cite{PhysRevB.86.134528}.
Additionally, a monochromatic optical field, promotes electrons
from the valence band (VB) (labeled as 3) to the CB (labeled as $1$) of QDA.
We consider only the CB of QDB (labeled as 2), as we have
assumed a large mismatch between the energies of the optical field and the band gap.
The QDs are also coupled to normal leads, indicated as $R$ and $L$ in Fig.\ref{fig:figure1}(a),
that operate as source or drain of particles~\cite{hiltscher2011}.

Based on Refs. [\onlinecite{PhysRevLett.120.087701}] and [\onlinecite{fujita2015single}],
we can write the Hamiltonian accounting for the spin degree of freedom as,
\begin{eqnarray}
\label{eq:fullH_wspin}
H_{\mathrm{full}}&=&\sum^{3}_{i=1}\sum_{\sigma=\uparrow,\downarrow}\varepsilon^0_{i\sigma}d^{\dagger}_{i\sigma}d_{i\sigma}
+\gamma_S\left(d^{\dagger}_{1\uparrow}d^{\dagger}_{2\downarrow}-d^{\dagger}_{1\downarrow}d^{\dagger}_{2\uparrow}\right)\nonumber\\
&&+\Omega_{-}e^{-i\omega t}d^{\dagger}_{1\uparrow}d_{3\uparrow}+\Omega_{+}e^{-i\omega t}d^{\dagger}_{1\downarrow}d_{3\downarrow}+\mathrm{H.c.},
\end{eqnarray}
where $d_{i\sigma}$ $(d_{i\sigma}^{\dagger})$ annihilates (creates) one electron with energy $\varepsilon^0_{i\sigma}$ and spin $\sigma$ in the $i$-th level.
The second term accounts for the CAR process with strength $\gamma_S$.
Terms with $\Omega_{\pm}$ provides the optical coupling of a circularly polarized light in QDA, that selects a specific spin component ($\Omega_{-}$ for spin $\uparrow$ and $\Omega_{+}$ for spin $\downarrow$)~\cite{fujita2015single,Villas07}. Here $\omega$ is the frequency of the optical field.

In principle, the energies $\varepsilon^0_{i\sigma}$ has spin degeneracy, $\varepsilon^0_{i\uparrow}=\varepsilon^0_{i\downarrow}$ ($i=1,2,3$). In our specific setup, we assume the use of local magnetic fields in the QDs, which lifts the energy levels degeneracy due to the Zeeman interaction. In this simplified model, for instance,  $\varepsilon_{1\uparrow}=\varepsilon^0_{1\uparrow}-|\varepsilon_Z|$
and $\varepsilon_{1\downarrow}=\varepsilon^0_{1\downarrow}+|\varepsilon_Z|$ for QDA, while
$\varepsilon_{2\uparrow}=\varepsilon^0_{2\uparrow}+|\varepsilon_Z|$ and $\varepsilon_{2\downarrow}=\varepsilon^0_{2\downarrow}-|\varepsilon_Z|$
for QDB, by considering appropriate orientations and strengths of local magnetic fields on each quantum dot, and calling $|\varepsilon_Z|$ as the Zeeman energy.
In addition, by tuning $\varepsilon_{1\sigma}^0$ and $\varepsilon_{2\sigma}^0$
with gate voltages, one can set $\varepsilon_{1\uparrow}$ and $\varepsilon_{2\downarrow}$ in resonance with the Fermi level of the superconductor lead.
Moreover, by optically selecting one spin component (e.g. $\Omega_{+}=0$ and $\Omega_{-} \neq 0$),
the specific spin configuration $\ket{1{\uparrow};2{\downarrow};3{\uparrow}}$ turns out to be more accessible
than other possible alignments. In a similar way, one could tune the parameters
to select the spin configuration $\ket{1{\downarrow};2{\uparrow};3{\downarrow}}$. An alternative setup consists on the action of magnetic field over the whole system (quantum dots and leads) in the Coulomb blockade regime, so it acts as a spin filter~\cite{PhysRevLett.85.1962}.

Once the spin degree of freedom is fixed, we can work with the effective spinless Hamiltonian,
\begin{equation}
\label{eq:Hspinless}
H_{\mathrm{sl}}=\sum_{i=1}^{3}\varepsilon_i d^{\dg}_i d_i+ \Omega e^{-i\w t} d^{\dg}_1 d_3+\gamma_{S} d^{\dg}_1 d^{\dg}_2 + \mathrm{H.c.},
\end{equation}
where the first term is the free Hamiltonian for the QDs, and $d_i$ ($d_i^\dagger$)
annihilates (creates) one electron in level $i$ with energy $\varepsilon_{i}$.
Local electric fields can be used to manipulate the difference between the energies of the conduction levels,
defined as $\delta E=\varepsilon_2-\varepsilon_1$. This model shares similarities with those found in quantum optics and atomic physics, systems with strong spatial confinement of electronic states~\cite{Scullybook,Muller12}. The CAR process plays a similar role as the tunneling in double QDs with excitons~\cite{Holmes13,Borges13,Laucht09,Villas05,Villas04}.
In the Appendix, we compare the eigenvalues and eigenvectors for both $H_{\mathrm{full}}$ and $H_{\mathrm{sl}}$. By appropriately tunning the physical parameters, a subset of eigenvalues and eigenvectors of  $H_{\mathrm{full}}$ matches to the ones found for $H_{\mathrm{sl}}$, corresponding to the relevant three-dimensional subspaces, as described in the section below.

\section{Effective three-level systems and formation of ATD}
\label{closedsystem}

Let us consider initially a closed quantum system ($\Gamma_2=\Gamma_3=0$) to focus on the action of the superconductor lead as a Cooper pair beam splitter (CPBS)~\cite{hofstetter2009,herrmann2010,PhysRevB.96.195409}. Assuming the computational basis composed by states of the form $\ket{n_1n_2n_3}$, with $n_i=1(0)$ indicating occupation (vacancy) of one electron in level $i$, there are two possible mechanisms for the formation of ATD, as illustrated in Fig.~\ref{fig:figure1}(b)and ~\ref{fig:figure1}(c). The first one [Fig.~\ref{fig:figure1}(b)] corresponds to a subspace spanned by states with an $\emph{even}$ number of particles,
i.e., $\left\{\ket{011},\ket{000},\ket{110}\right\}$. Two coupling mechanisms take place within this subspace: the optical
transition $\Omega \exp[i \omega t] \ket{011}\bra{110}+H.c.$ and the CAR process $\gamma_S\ket{110}\bra{000}+H.c.$
These two combined effects turn into a three level system in a $\Lambda$ configuration if $\delta E>0$.
The second one [Fig.~\ref{fig:figure1}(c)] corresponds to the subspace
spanned by $\left\{\ket{001},\ket{111},\ket{100}\right\}$,
with an $\emph{odd}$ number of particles. Here the states $\ket{001}$ and $\ket{100}$ are optically coupled, while
the states $\ket{001}$ and $\ket{111}$ couples via CAR, in a $V$ configuration if $\delta E>0$.
Both subspace can be experimentally accessed via suitable choice of states initialization.

Figures~\ref{fig:figure1}(d) and \ref{fig:figure1}(e) show the occupations $n_i=\mathrm{Tr}[d_i^\dagger d_i \rho_S(t)]$ ($i=2,3$),
as a function of time and $\delta E$ for a decoherence free system.
Here the system is initialized at $\rho_\mathcal{S}(0)=\ket{001}\bra{001}$.
The appearance of fast Rabi oscillations on $n_3$ [Fig.\ref{fig:figure1}(d)],
with time scale $t\sim 1/\Omega$ is a characteristic of a
two-level system under the action of optical fields.\cite{Zrenner2002}
More interestingly, in Fig.\ref{fig:figure1}(e) we see an enhancement of the population $n_2$ for two specific energy values, $\delta E=\pm\Omega$.
Those energies are characteristic of the ATD.

\begin{figure}[tbh]
\centering\includegraphics[width=1\linewidth]{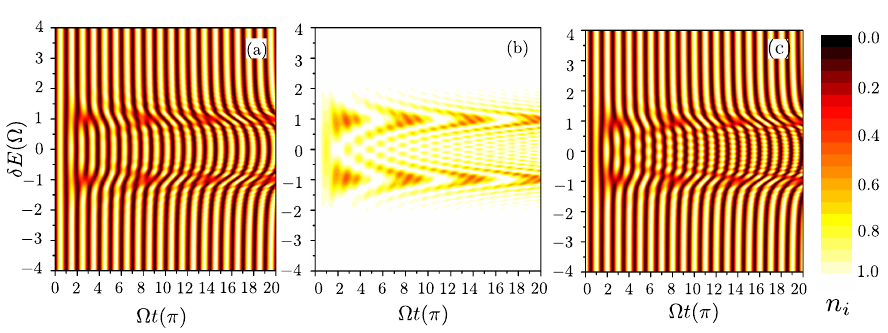}
\caption{(a)-(c) Closed dynamics of the quantum system ($\Gamma_2=\Gamma_3=0$),
as function of time and $\delta E=\varepsilon_2-\varepsilon_1$, considering the initial state $\ket{011}$
($\Lambda$ configuration): (a) occupations $n_1$ of the conduction band of QDA,
(b) occupation $n_2$ of the conduction band of QDB, and
(c) occupation $n_3$ of the valence band of QDA.  Physical parameters are set as $\varepsilon_1=0$, $\omega=\varepsilon_1-\varepsilon_3$
and $\gamma_S=\Omega/4$.}
\label{fig:fig2_si}
\end{figure}
In Fig. \ref{fig:fig2_si} we show the time evolution of the populations $n_1$, $n_2$ and $n_3$,
when the system is initialized at $\rho_\mathcal{S}(0)=\ket{011}\bra{011}$. Similar features
to those found in Figs.~\ref{fig:figure1}(d) and \ref{fig:figure1}(e) are observed here.
The main contrast is found for $n_2$, Fig. \ref{fig:fig2_si}(b): while in Fig.~\ref{fig:figure1}(e)
the Autler-Townes doublet is manifested via two peaks of $n_2$ at $\delta E=\pm \Omega$,
here the Autler-Townes doublet appears as two dips, as a result
of the initialization $n_2=1$.

\section{Dynamics of the open quantum system}
\label{sec:dynopen}
Once we have demonstrated the conditions for the appearance of the ATD in the closed system, we proceed to include the action of the reservoirs as considered in Fig.~\ref{fig:figure1}(a). The Hamiltonian including  reservoir terms would read as,
\begin{equation}
\label{eq:Hinit}
H=H_{\mathrm{sl}}+\sum_{\eta,k_{\eta}} \varepsilon_{k_{\eta}} c^{\dg}_{k_{\eta}} c_{k_{\eta}} + \sum_{k_3}V_{3} d^{\dg}_3 c_{k_3}+\sum_{k_2}V_{2} d^{\dg}_2 c_{k_2} + \mathrm{H.c.}
\end{equation}
The second term in Eq.~(\ref{eq:Hinit}) is the free Hamiltonian for the reservoirs, where the operators $c_{k_\eta}$ ($c_{k_\eta}^\dagger$) annihilates (creates) electrons with momentum $k$ in reservoir $\eta$.
The last two terms describe the electronic tunneling between the QDs and reservoirs,
with momentum independent strengths $V_i$ ($i=2, 3$). Here $\eta=2(3)$ for the right (left) reservoir.
The quantum dynamics and the electronic transport are obtained by solving a differential
equation for the reduced density matrix $\rho_S(t)$ of the quantum dots system.
The first step is to perform the
unitary transformation $U(t)=\exp{[i \omega t(d^{\dg}_1 d_1-d^{\dg}_2 d_2-d^{\dg}_3 d_3-\sum_{\eta,k_{\eta}}c^{\dg}_{k_{\eta}} c_{k_{\eta}})/2]}$,
which drops the time-dependent exponential from the optical pumping. The transformed Hamiltonian would read as $H'=H_0+V$, where
\begin{eqnarray}
\label{eq:Htrans}
H_0&=&\sum_{i=1}^{3}\widetilde{\ve}_i d^{\dg}_i d_i+\sum_{\eta,k_{\eta}} \widetilde{\ve}_{k_{\eta}} c^{\dg}_{k_{\eta}} c_{k_{\eta}}+\Omega d^{\dg}_1 d_3+\gamma_S d^{\dg}_1 d^{\dg}_2+ \mathrm{H.c.}\nonumber\\
V&=&\sum_{k_3}V_{3}d^{\dg}_3 c_{k_3}+\sum_{k_2}V_{2} d^{\dg}_2 c_{k_2}+ \mathrm{H.c.},
\end{eqnarray}
with $\widetilde{\ve}_1=\ve_1-\omega/2$, $\widetilde{\ve}_{2(3)}=\ve_{2(3)}+\omega/2$ and $ \widetilde{\ve}_{k_{\eta}}=\ve_{k_{\eta}}+\omega/2$. The evolution of the density matrix $\rho(t)$ for the full system (dots and reservoirs) is given by the Von Neumann equation, $\dot{\rho}(t)=-i [H',\rho(t)]$ ($\hbar=1$). We first write $\hat{\rho}(t)=e^{i H_0 t} \rho(t) e^{-i H_0 t}$, where the \textit{hat} symbol over the operators stands for the interaction picture. The exact solution for the dynamics of the system is given by $\dot{\hat{\rho}}(t)= \mathcal{L}(t) \hat{\rho}_0 + \int_{0}^t dt_1 \mathcal{L}(t) \mathcal{L}(t_1)\hat{\rho}(t_1)$ where $\mathcal{L}(t)$ is the Liouvillian superoperator, $\mathcal{L}(t)\hat{\rho}(t_1)=-i [\hat{V}(t),\hat{\rho}(t_1)]$ and $\hat{V}(t)$ is the dots-to-reservoirs coupling in the interaction picture.

At this point, we use the Born approximation $\hat{\rho}(t)=\hat{\rho}_S(t) \otimes \hat{\rho}_L \otimes \hat{\rho}_R$,
where $\hat{\rho}_S(t)=\mathrm{Tr}_{L+R}[\hat{\rho}(t)]$ is the reduced matrix after taking
the partial trace over the reservoirs degrees of freedom.
The quantities $\hat{\rho}_L$ and $\hat{\rho}_R$ are the density matrices for the left and
right reservoirs.
Within this calculation, we arrive in a integro-differential equation that describes the dynamics of the reduced density matrix $\hat{\rho}_S(t)$:
\begin{eqnarray}
\label{eq:rhointeraction}
&&\dot{\hat{\rho}}_S(t)=-i \int_0^t dt_1 \sum_{i,j} \left[g^{>}_{ij}(t,t_1)\hat{d}_i^\dagger (t) \hat{d}_j (t_1) \hat{\rho}_S(t_1)\right.\\
&&\left.- g^{>}_{ji}(t_1,t)\hat{d}_i(t) \hat{\rho}_S(t_1) \hat{d}_j^\dagger (t_1)- g^{<}_{ji}(t_1,t)\hat{d}_i(t) \hat{d}_j^\dagger(t_1) \hat{\rho}_S(t_1)\right.\nonumber \\
&&\left.+g_{ij}^{<} (t,t_1)\hat{d}_i^\dagger(t) \hat{\rho}_S(t_1)\hat{d}_j(t_1)\right] + \mathrm{H.c.}\nonumber,
\end{eqnarray}
where each term contains the first-order correlation functions for the free-electrons on reservoirs defined as $g^<_{ij}(t,t{'}) = \delta_{ij} |V_i|^2 \sum_{k_i}i\langle \hat{c}_{k_i}^\dagger(t{'}) \hat{c}_{k_i}(t)\rangle$ and
$g^>_{ij}(t,t{'}) = \delta_{ij} |V_i|^2  \sum_{k_i}  (-i) \langle \hat{c}_{k_i}(t) \hat{c}_{k_i}^\dagger(t{'}) \rangle$.
In the wideband approximation~\cite{Jauhobook}, they take the form
\begin{eqnarray}
 \label{eq:greens}
 g^<_{ij}(t,t_1) &=& i \delta_{ij} 2\pi \mathcal{D}_i |V_i|^2 f_i \delta(t-t_1)\\
 g^>_{ij}(t,t_1) &=& -i \delta_{ij} 2\pi \mathcal{D}_i |V_i|^2 (1-f_i) \delta(t-t_1),\nonumber
\end{eqnarray}
where the function $f_i$ is the Fermi function and $\mathcal{D}_i$ is a constant density of states for $i$-th reservoir.
Fermi functions take the values $f_i=0$ ($f_i=1$), if the reservoir is a drain (source) of particles.
Using Eq.~(\ref{eq:greens}) into Eq.~(\ref{eq:rhointeraction}) and integrating over time, we obtain
\begin{eqnarray}
&&\dot{\hat{\rho}}_S(t)=\frac{1}{2} \sum_{i} \Gamma_i \left\{-(1-f_i)\left[\hat{d}_i^\dagger (t) \hat{d}_i (t) \hat{\rho}_S(t)\right.\right.\nonumber \\
&&\left.\left.-\hat{d}_i(t) \hat{\rho}_S(t) \hat{d}_i^\dagger (t)\right]-f_i\left[\hat{d}_i(t)\hat{d}_i^\dagger(t) \hat{\rho}_S(t)\right.\right.\nonumber \\
&&\left.\left.-\hat{d}_i^\dagger(t) \hat{\rho}_S(t)\hat{d}_i(t)\right]+\mathrm{H.c.}\right\},
\end{eqnarray}
with $\Gamma_i=2\pi \mathcal{D}_i |V_i|^2$ being the strength of the coupling to the reservoirs.
In the Schr\"odinger picture, we find the Lindblad equation~\cite{Lindblad76},
\begin{equation}
\label{eq:Lindblad}
\dot{\rho}_S=-i [H_S,\rho_S]-\frac{1}{2} \sum_{i} \Gamma_i\left[f_i\mathcal{L}^{+}_i+(1-f_i)\mathcal{L}^{-}_i\right],
\end{equation}
where $H_S$ is the system Hamiltonian, without the reservoirs terms in Eq.~(\ref{eq:Htrans}).
The dissipative terms describing the action of the reservoirs (source or drain) of electrons
are given by $\mathcal{L}^{+}_i=d_i d_i^\dagger \rho_S + \rho_S d_i d_i^\dagger - 2 d_i^\dagger \rho_S d_i$,
and $\mathcal{L}^{-}_i=d_i^\dagger d_i \rho_S + \rho_S d_i^\dagger d_i - 2 d_i \rho_S d_i^\dagger$.

Performing a vectorization procedure of the density matrix, $\vec{\rho}_S=\mathbf{vec}[\rho_S]$,
Eq. (\ref{eq:Lindblad}) takes the following form~\cite{souza2017}
\begin{equation}
\label{eq:vecrhos}
\vec{\rho}_S(t)=e^{-i \mathcal{M} t} \vec{\rho}_S(0),
\end{equation}
where $\mathcal{M}$ is a matrix with dimension $I^{\otimes 3}\otimes I^{\otimes 3}$, for $I$ being the $2D$ identity matrix. This matrix is defined as $\mathcal{M}=\mathcal{M}_0-i \mathbf{\Gamma}/2$, where $\mathcal{M}_0=I^{\otimes 3}\otimes H_S - H_S^T \otimes I^{\otimes 3}$ describes the system (superscript $T$ meaning matrix transposition) and $\mathbf{\Gamma}$ contains the effect of reservoir, defined as $\mathbf{\Gamma}=\sum_i \Gamma_i f_i \mathbf{L}^{+}_i+\Gamma_i
  (1-f_i)\mathbf{L}^{-}_i$, with $\mathbf{L}^{+}_i=I^{\otimes 3} \otimes d_i
  d_i^\dagger +(d_i d_i^\dagger)^T \otimes I^{\otimes 3} -2 d_i \otimes d_i$
  and $\mathbf{L}^{-}_i=I^{\otimes 3} \otimes d_i^\dagger d_i +(d_i^\dagger
  d_i)^T \otimes I^{\otimes 3} -2 d_i^\dagger \otimes
  d_i^\dagger$.~\cite{souza2017}

In order to solve Eq.~(\ref{eq:vecrhos}), by numerical procedures,
we write operators $d_i$ in terms of the Jordan-Wigner transformation~\cite{Schallerbook,souza2017}
as  $d_1=\sigma_{-} \otimes \sigma_z \otimes \sigma_z$,
$d_2=I \otimes \sigma_{-} \otimes \sigma_z$ and $d_3=I \otimes I \otimes \sigma_{-}$.
This procedure automatically writes any state or operators associated with the system of
quantum dots in the basis given
by $\{|111\rangle,|110\rangle,|101\rangle,|100\rangle,|011\rangle,|010\rangle,|001\rangle,|000\rangle\}$,
These eight elements permit the description of a nonequilibrium scenario where electrons can flow \emph{in} and \emph{out} of the quantum dots.

\section{Signatures of ATD on photocurrent}
\label{sec:photocurrent}

\begin{figure}[b]
\centering\includegraphics[width=0.9\linewidth]{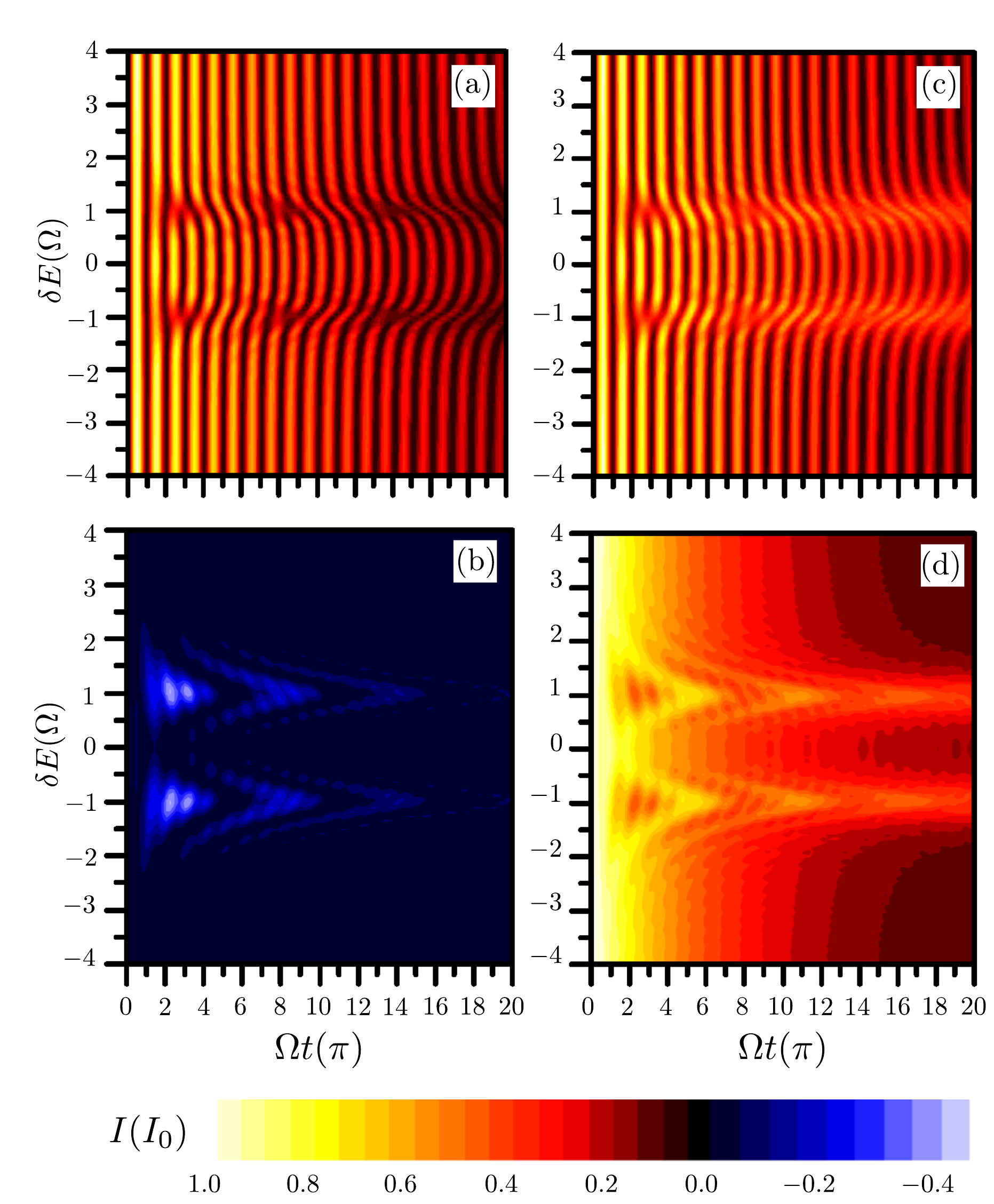}
\caption{The currents $I_L$, panels (a) and (c), and $I_R$, panels (b) and (d) (in unities of $I_0$),
as functions of time and $\delta E$, and the initial
condition is set as $\rho_S(0)=\ket{001}\bra{001}$. In panels (a) and (b), $f_2=0$ so the reservoir $R$ acts as a drain of electrons. In panels (c) and (d), the reservoir becomes a source of electrons with $f_2=1$.
Physical parameters: $\varepsilon_1=0$, $\omega=\varepsilon_1-\varepsilon_3$, $\Gamma=\gamma_S/5$ and $\gamma_{S}=\Omega/4$.}
\label{fig:figure2}
\end{figure}

To obtain the electronic currents through $L$ and $R$ reservoirs,
we use rate equations~\cite{Souza07}, i.e., $I_{R}=I_0\left[f_2P^0_{2}-(1-f_2)P^1_{2}\right]$ and $I_{L}=I_0 f_3 P^0_{3}$,
where $I_0=e\Gamma/\hbar$, with $\Gamma_2=\Gamma_3=\Gamma$. The occupation probabilities are calculated according to
$P^l_2=\sum_{n,m=0}^{1}\bra{nlm}\rho_{S}\ket{nlm}$, where $l=0$ ($1$) if CB on QDB is empty (full), and
$P^0_3=\sum_{n,l=0}^{1}\bra{nl0}\rho_{S}\ket{nl0}$, for an empty VB in QDA.
Fig. \ref{fig:figure2} shows the left ($I_L$) and the right ($I_R$)
currents for the cases where the right reservoir acts as a (i) drain $f_2=0$
[Figs.~\ref{fig:figure2}(a)-(b)], or as a (ii) source $f_2=1$ [Figs.~\ref{fig:figure2}(c)-(d)].
Concerning the first case, Fig.~\ref{fig:figure2}(a) shows that $I_L$ has a positive value,
meaning that the electrons flow from the reservoir $L$ into QDA, as they are being photoexcited.
The signature of the action of optical pumping is an oscillation at short times that matches with the
Rabi oscillations shown in Fig.~\ref{fig:figure1}(d). As time increases, the incoherent coupling
between the system and the reservoirs causes the attenuation of these oscillations, with the current being suppressed for increasing times.
Interestingly, the current in the right electrode shows negative values for the condition
$\delta E=\pm\Omega$ as seen in Fig.~\ref{fig:figure2}(b), although, both $I_R$ and $I_L$ go to zero as time evolves.
The explanation for such a behavior is as follows: the $R$ electrode operates as a drain of electrons, so
whenever an electron is created in QDB via CAR, it has a finite probability to be drained into the right lead,
generating an outgoing probability current. As soon as the electron leaves QDB, its pairing electron in the CB of QDA stays locked, thus forbidding further optical transitions or CAR process. This fact results on a vanishing photocurrent at the stationary regime, so we can assert that this configuration permits the detection of ATD from current measurements only at the transient time scale.

The previous situation changes when $f_2=1$, as shown in Fig.~\ref{fig:figure2}(c)-(d).
Again, we still observe the signatures of Rabi oscillations on both currents, however,
two main differences can be noticed: (i) $I_R$ takes positive values only,
due to the fact that electrode $R$ acts as a source, (ii) high values of current, even at long times,
are predicted at the condition of ATD, $\delta E=\pm\Omega$. These differences came from the fact that,
while QDB is populated by electrons coming from the right lead, electrons are optically pumped in QDA.
At this point, CAR process annihilates both electrons in the dots,
opening the possibility of further injection and optical excitation of
electrons in the dots.
Then the sequence repeats again, resulting in a finite current in the stationary regime.

\begin{figure}[tbh]
\centering\includegraphics[width=1\linewidth]{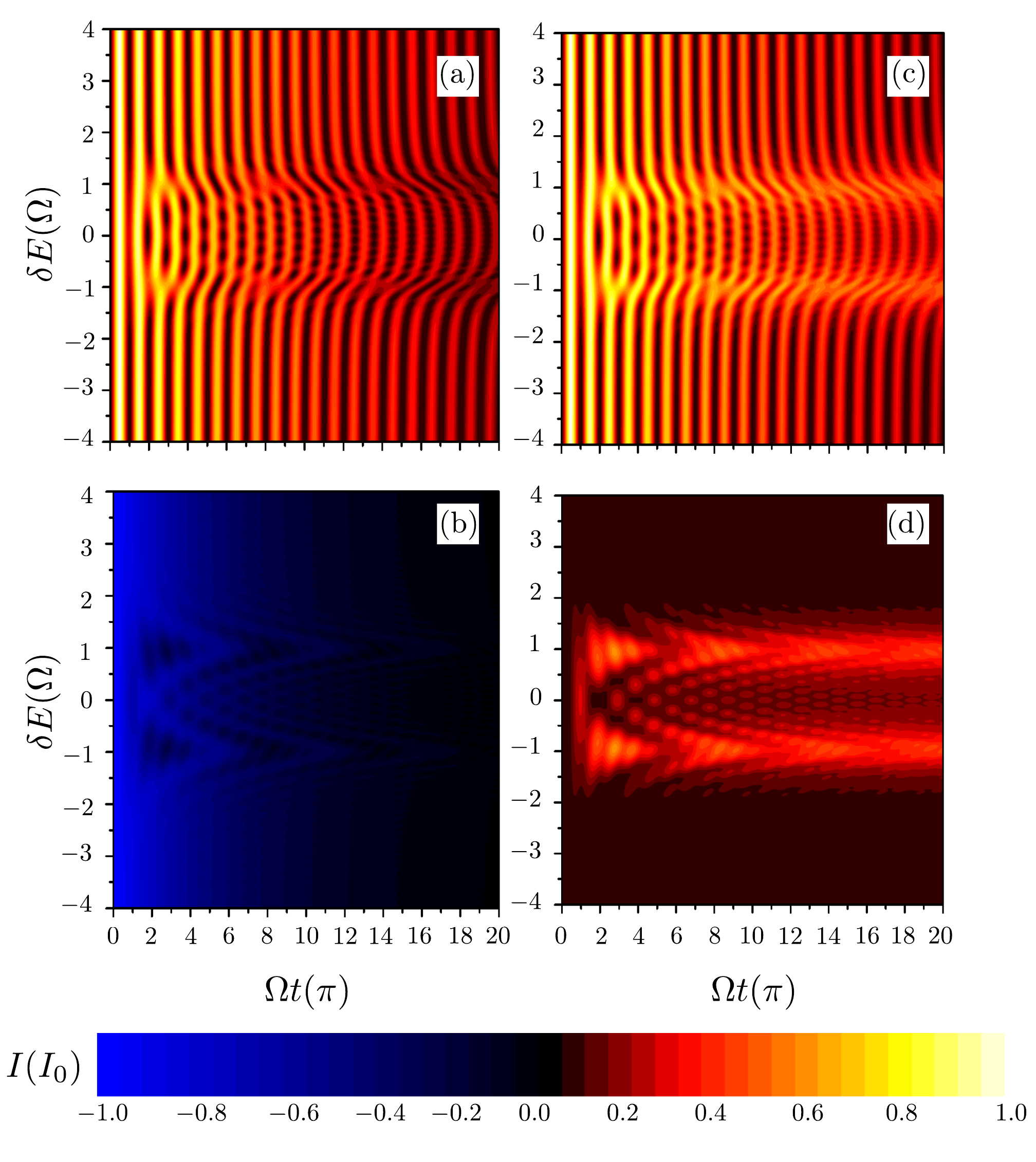}
\caption{The currents $I_L$, panels (a) and (c), and $I_R$, panels (b) and (d), in unities of $I_0$,
as functions of time and $\delta E$ considering $\Gamma=\gamma_{S}/5$ and the initial
condition $\rho_S(0)=\ket{011}\bra{011}$, which is part of the $\Lambda$ configuration.
In panels (a)-(b), $f_2=0$ so the reservoir $R$ acts as a drain of electrons.
In panels (c)-(d), the reservoir becomes a source of electrons with $f_2=1$.
Physical parameters: $\varepsilon_1=0$, $\omega=\varepsilon_1-\varepsilon_3$, and $\gamma_{S}=\Omega/4$.}
\label{fig:fig3_si}
\end{figure}
The same features can be seen if the system is initialized at $\rho_S(0)=\ket{011}\bra{011}$. In Fig.\ref{fig:fig3_si} we show the currents $I_L$ and $I_R$ as a function of time
and $\delta E$. The major contrast between these results to those shown in Fig.\ref{fig:figure2} is observed at transient time scales. In Fig. \ref{fig:fig3_si}(b)
the currents $I_R$ is predominantly negative for short time scales. This behavior is noticed because
QDB is initially populated by a single electron, so this is observed as a finite current probability
for the electron in QDB to tunnel to the right reservoir ($f_2=0$), at short times.
In contrast, the current $I_R$ seen in Fig.\ref{fig:fig3_si}(d) ($f_2=1$) starts at zero.
Since QDB is initially occupied by a single
electron, this particle should be drained into the superconductor lead before a current takes
place from the right reservoir ($f_2=1$) into QDB. The Autler-Townes doublet found in the stationary
regime remains the same regardless of initial state. Because this stationary current is significantly high at the ATD condition $\delta E=\pm\Omega$,
our results are a \emph{proof-of-principle} that this optical phenomena can be detected by performing current measurements.

\begin{figure}[tb]
\centering\includegraphics[width=0.9\linewidth]{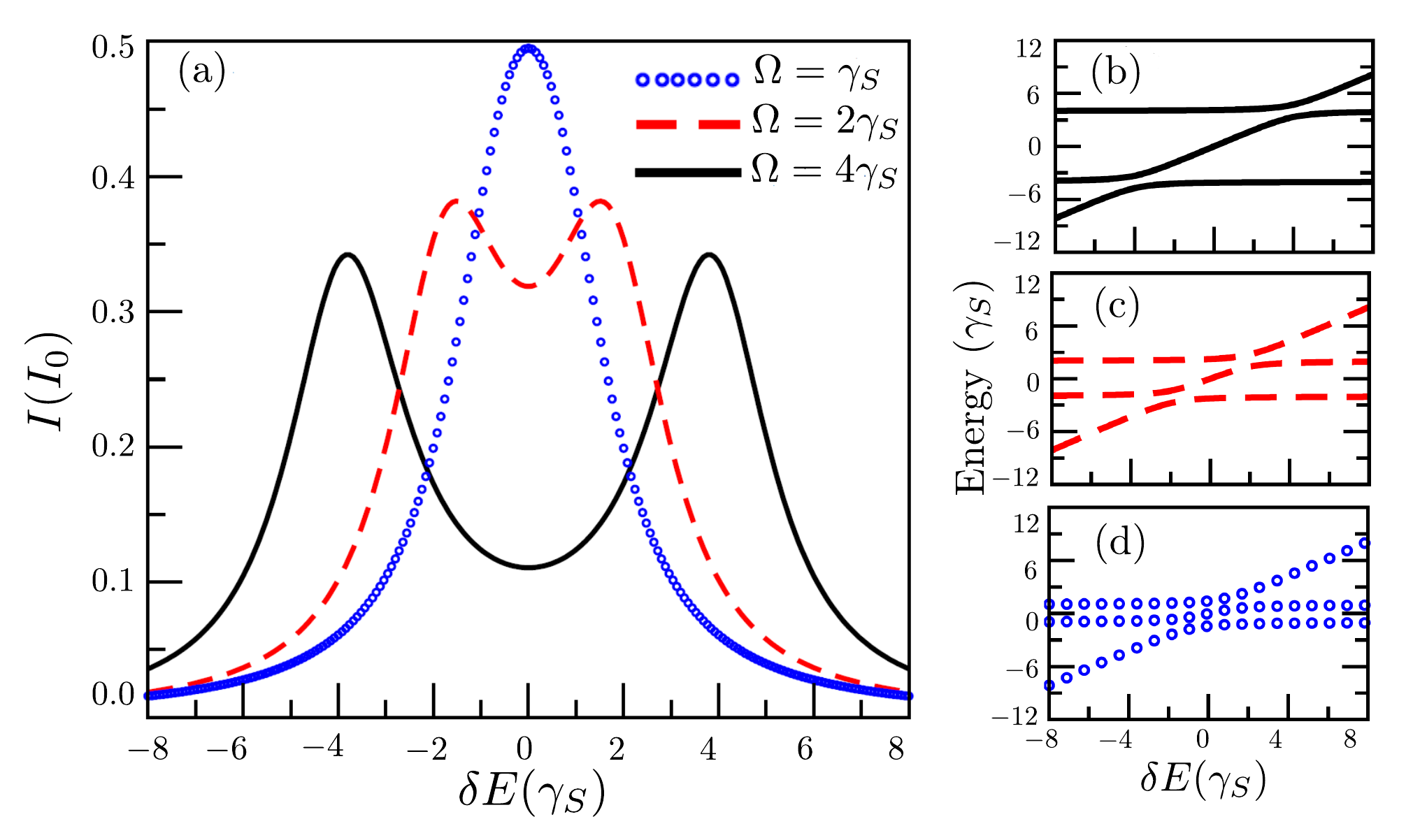}
\caption{(a) Autler-Townes doublet observed in the electric current considering $\Gamma=\gamma_{S}/5$. (b)-(d): The eigenvalues of the effective three-level system
for $\Omega=4\gamma_{S}$ (black solid line), $2\gamma_{S}$ (red dashed line),  and $\gamma_{S}$ (blue open circles), respectively.}
\label{fig:figure4}
\end{figure}

To conclude our discussion, we now show the behavior of the characteristic ATD profile in Fig.~\ref{fig:figure4}(a),
as imprinted in the stationary current $I=I_R=I_L$, as a function of $\delta E$ for different values of $\Omega$.
From this panel, it is clear to notice two resolved peaks at $\delta E\approx\pm\Omega$, for $\Omega/\gamma_{S} > 1$,
while for $\Omega/\gamma_{S}\rightarrow 1$, these two peaks merge into a single one.
This behavior resembles to the characteristic profile of the luminescence spectrum
for an excitonic system, when laser intensity is varied~\cite{kamada2001}.
In order to clarify the appearance of the peaks on the current, in Fig.~\ref{fig:figure4}(b)-(d),
we show the eigenvalues associated to the three-level subspace formed by $\{\ket{001},\ket{111},\ket{100}\}$
for different values of $\Omega=4\gamma_{S}$, $2\gamma_{S},$ and $\gamma_{S}$ (eigenvalues were shifted to appear around zero).
The eigenvalues present anticrossings around the Rabi energy,
revealing a strong coupling between the levels of the quantum dots when $\delta E\approx\pm\Omega$.
This coupling opens up a transmission channel in the system that allows the flow of a stationary current.
The values of optical pumping strength close to $\gamma_{S}$ yields to a low resolution of anticrossings,
as shown in Fig.~\ref{fig:figure4} (d) for $\Omega=\gamma_{S}$, which in transport appears as the
single peak (blue open circles) in Fig.~\ref{fig:figure4}(a).
\section{Summary}
\label{sec:summary}
We present a proposal for the detection of an optical phenomenon through measurements
of quantum transport in a nonequilibrium system. We first describe the formation process of an
Autler-Townes doublet on quantum dots coupled with a superconductor lead, which results
from the combination of the action of an optical pumping and crossed Andreev reflection.
Calculations performed with density matrix formalism shows that signatures of the formation of
this doublet can be found on transport measurements, even in a stationary regime,
with the appropriate parameter conditions.

\acknowledgments
This work was supported by CNPq (grant 307464/2015-6), FAPEMIG, and the Brazilian National Institute of Science and Technology of Quantum Information (INCT-IQ).

\appendix
\section{Comparison between the spinfull and spinless model.}
\label{app:spectrum}
\begin{figure}[tb]
\centering\includegraphics[width=0.9\linewidth]{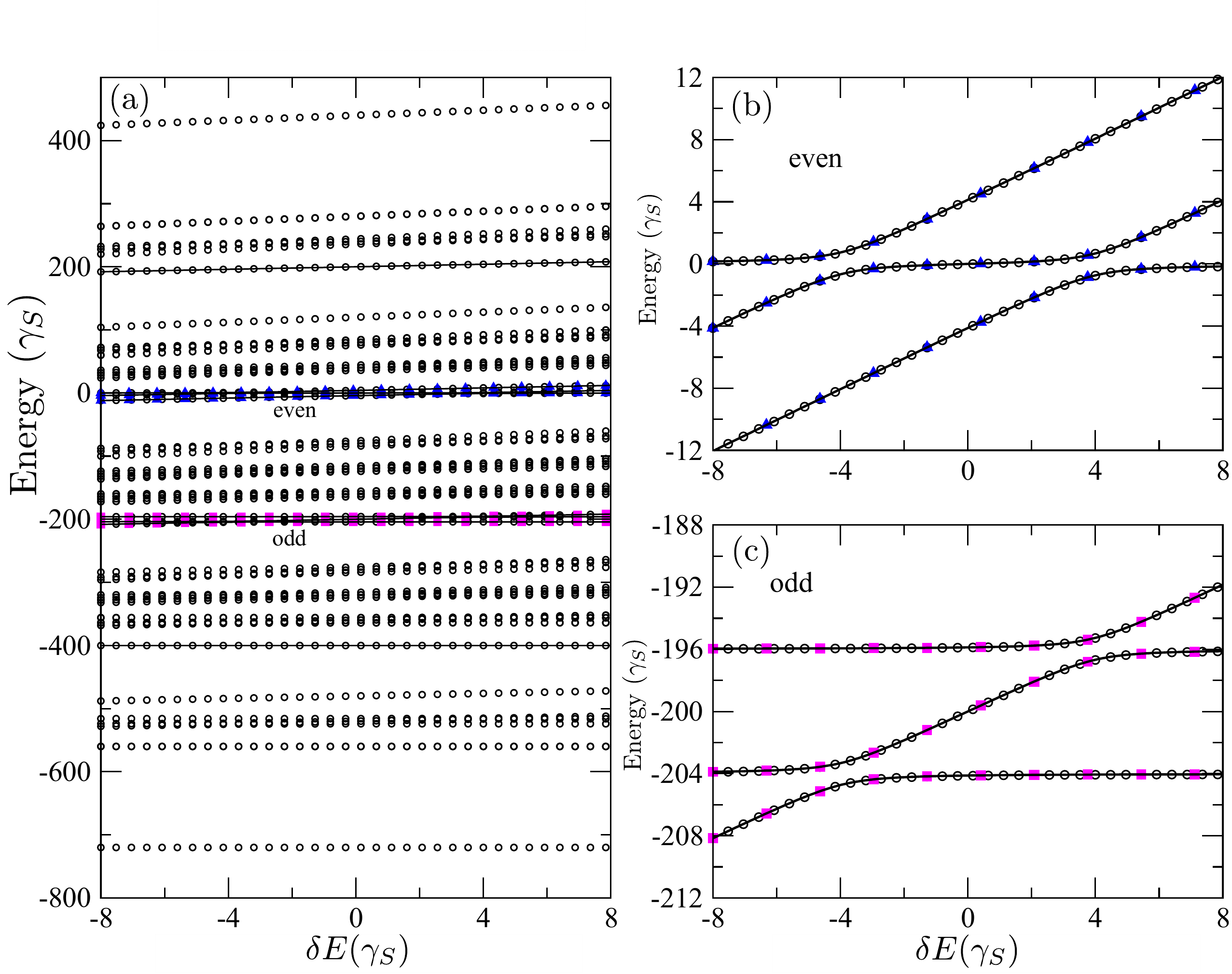}
\caption{Comparison between eigenvalues of Hamiltonian (\ref{eq:fullH_wspin}) and the spinless Hamiltonian (\ref{eq:Hspinless}): (a) energy spectrum of $H_{\mathrm{full}}$ (open circles) and $H_{\mathrm{sl}}$ (solid black line) as function of $\delta E$. (b) Zoom in for the even branch and for the odd branch in panel (c). The blue-triangles and the pink-squares represents the energy eigenvalues for the even and odd family of states for the spinless model, respectively. The parameters used are $|\varepsilon_Z|=5\Omega_{{-}}$, and $\omega=100\Omega_{-}$, with $\Omega_{-}=4\gamma_S$ and $\Omega_{+}=0$.}
\label{fig:fig1_si}
\end{figure}

In this appendix, we compare the spectrum of the effective spinless model, Eq.(\ref{eq:Hspinless}) to the spinfull Hamiltonian $H_{\mathrm{full}}$, given by Eq.(\ref{eq:fullH_wspin}). To do so, we consider a basis composed of states in the form $\ket{n_{1\uparrow},n_{1\downarrow};n_{2\uparrow},n_{2\downarrow};n_{3\uparrow},n_{3\downarrow}}$,
where $n_{i\sigma}$ assumes the value $1$ ($0$), indicating that state $i$ with spin projection $\sigma$ is occupied (empty).
Here, we follow again the convention $i=1$ for CB of QDA, $i=2$ for CB of QDB and $i=3$ for VB of QDA.
This basis includes all possible combinations, with $2^6=64$ states,
ranging from $\ket{1_{\uparrow}1_{\downarrow};1_{\uparrow}1_{\downarrow};1_{\uparrow}1_{\downarrow}}$
to $\ket{0_{\uparrow}0_{\downarrow};0_{\uparrow}0_{\downarrow};0_{\uparrow}0_{\downarrow}}$. To drop the exponential factor $e^{\pm i\omega t}$, we apply the unitary transformation
$U_f(t)=\exp\left[\frac{i\omega t}{2}\left(d^{\dagger}_{1\uparrow}d_{1\uparrow}+d^{\dagger}_{1\downarrow}d_{1\downarrow}
-\sum^{3}_{i=2}\sum_{\sigma=\uparrow,\downarrow}d^{\dagger}_{i\sigma}d_{i\sigma}\right)\right]$,
over the time-dependent Hamiltonian $H_{\mathrm{full}}$, Eq. \ref{eq:fullH_wspin},
and then use a numerical procedure of diagonalization to obtain the dressed eigenvalues and eigenstates.
The eigenvalues of the transformed spinfull Hamiltonian is shown as open circles in Fig.~\ref{fig:fig1_si}(a),
as we vary the energy difference $\delta E=\varepsilon_{2\downarrow}-\varepsilon_{1\uparrow}$.
The parameters used here are set as $|\varepsilon_Z|=5\Omega_{{-}}$, $\Omega_{-}=4\gamma_S$ and $\Omega_{+}=0$, and
the optical frequency $\omega$ is assumed to be much larger than other energy scales of the system, $\omega=100\Omega_{-}$.
Additionally, the levels $\varepsilon_{2\downarrow}$ and $\varepsilon_{1\uparrow}$ are set around zero.

One can notice from Fig.~\ref{fig:fig1_si}(a) that the energy spectrum shows families of branches,
and inside each family the levels are spaced by $\sim |\varepsilon_Z|$,
while the energy shift between families is of the order of $\omega/2$, which results from
the unitary transformation. From this complete spectrum, we can identify two particular branches of three eigenvalues each.
For sake of clarity, we have labeled these branches in Fig. \ref{fig:fig1_si} as $\emph{even}$ and $\emph{odd}$, where
the first one is found around $E=0.0\gamma_S$ while the second has energies around $E=-200\gamma_S$.

In Fig.\ref{fig:fig1_si}(b) and Fig.(c)
are shown a zoom of these two specific groups, where we can notice two anticrossings
at $\delta E=\pm\Omega_{-}$. By checking the eigenstates,
we have verified that the eigenvalues around $E=0.0\gamma_S$ are combinations of the following states with even number of particles:
$\ket{0_{\uparrow},0_{\downarrow};0_{\uparrow},1_{\downarrow};1_{\uparrow},0_{\downarrow}}$,
$\ket{0_{\uparrow},0_{\downarrow};0_{\uparrow},0_{\downarrow};0_{\uparrow},0_{\downarrow}}$
and $\ket{1_{\uparrow},0_{\downarrow};0_{\uparrow},1_{\downarrow};0_{\uparrow},0_{\downarrow}}$.
However, the second branch with eigenvalues around $E=-200\gamma_S$ is associated with
$\ket{0_{\uparrow},0_{\downarrow};0_{\uparrow},0_{\downarrow};1_{\uparrow},0_{\downarrow}}$,
$\ket{1_{\uparrow},0_{\downarrow};0_{\uparrow},1_{\downarrow};1_{\uparrow},0_{\downarrow}}$ and
$\ket{1_{\uparrow},0_{\downarrow};0_{\uparrow},0_{\downarrow};0_{\uparrow},0_{\downarrow}}$,
all states with an odd number of particles. As the eigenvalues forming each branch are relatively
well separated in energy from the other eigenvalues,
we can expect that these two groups of three states each, the \textit{even} and \textit{odd} ones,
can behave as two independent three-level systems.

In principle, one could expect additional
transition processes that couple the subspaces spanned by the \emph{even}
and \emph{odd} states to additional states of the whole basis,
not accounted for in the simplified Hamiltonian $H_{\mathrm{sl}}$.
However, such processes are suppressed for the set of parameters adopted.
This becomes clear when we compare the eigenvalues of $H_{\mathrm{full}}$
to the ones of the spinless model $H_{\mathrm{sl}}$.
In Fig.~\ref{fig:fig1_si}(b)-(c) we also show the eigenvalues of $H_{\mathrm{sl}}$ as solid lines.
Notice that the eigenvalues for both $H_{\mathrm{full}}$ and $H_{\mathrm{sl}}$ are in perfect agreement.
Also, the eigenstates of $H_{\mathrm{sl}}$ corresponding to the \emph{even} subspace are linear
combinations of $\ket{011},\ket{000},\ket{110}$, while for the \emph{odd} subspace are linear combinations
of $\ket{001},\ket{111},\ket{100}$. We plot the eigenvalues of the two separate three level systems as blue-triangles for the even family and pink-squares for the odd family of state in Figs.\ref{fig:fig1_si}(a)-(c).

%

\end{document}